\documentclass[12pt]{iopart}
\usepackage{iopams}
\usepackage{epsfig}

\begin{document}
\title{$\Xi^{-}$ and $\bar{\Xi}^{+}$ production in Pb+Pb collisions at~40~$A$~\hspace{0.08cm}GeV at CERN SPS}
\author {Christine Meurer for the NA49 Collaboration 
\footnote[3]{For the author list see Friese V {\it These proceedings}}
}
\ead{Christine.Meurer@cern.ch}

\address{Gesellschaft f\"ur Schwerionenforschung, Planckstr.1, D-64291 Darmstadt, Germany}

\begin{abstract}
\noindent
First results on the production of $\Xi^{-}$ and $\bar{\Xi}^{+}$ hyperons in Pb+Pb interactions at 40~$A$\hspace{0.08cm}GeV are presented. The $\bar{\Xi}^{+}/\Xi^{-}$ ratio at midrapidity is studied as a function of collision centrality. The ratio shows no significant centrality dependence within statistical errors; it ranges from 0.07 to 0.15. The $\bar{\Xi}^{+}/\Xi^{-}$ ratio for central Pb+Pb collisions increases strongly with the collision energy.

\end{abstract}

\section{Introduction}
The measurement of multiple strange particles provides information on the hadronization features of strongly interacting matter produced in heavy ion collisions. If we assume that a deconfinement phase transition occurs in a heavy ion collision, we have to look for qualitative changes (onset phenomena) when varying external parameters like system size and energy density. Therefore we study the $\Xi^{-}$ and $\bar{\Xi}^{+}$ production as a function of collision centrality and beam energy. This contribution reports on the first results of centrality and energy dependence of the $\bar{\Xi}^{+}/\Xi^{-}$ ratio and compares it with predictions of the hadron-gas model \cite{bgs98}.

\section{Experimental Procedure}
\subsection{The NA49 Experiment}
\begin{figure}[h]
\begin{center}
\includegraphics[scale=0.45]{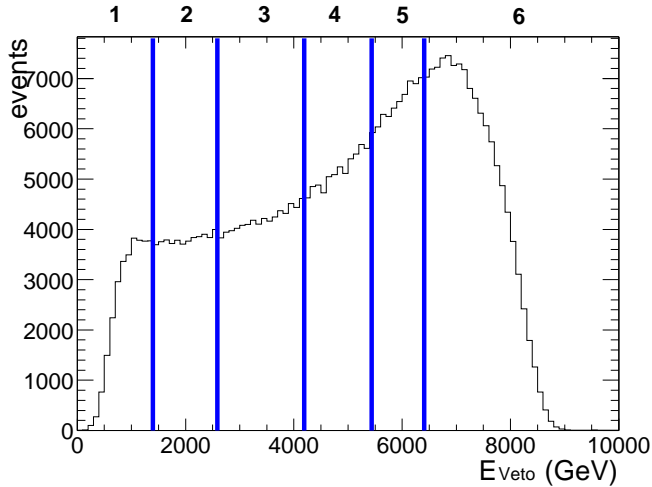}
\caption{\label{Eveto} Spectrum of the energy measured by VCAL for Pb+Pb collisions at 40~$A$~\hspace{0.08cm}GeV (minimum bias data)}
\end{center}
\end{figure}

The NA49 experiment \cite{NA49setup} is a large acceptance hadron spectrometer which consists of four TPCs. The two Vertex TPCs are located in a strong magnetic field. They allow the measurement of charge and momentum of the detected particles. The two Main TPCs are outside the magnetic field. They are used for particle identification via dE/dx measurement. Downstream of the TPCs a veto-calorimeter (VCAL) measures the energy of spectator nucleons from which the collision centrality can be inferred. In figure~\ref{Eveto} the distribution of the energy measured by VCAL for minimum bias data is shown. For the current $\Xi$ analysis we divide this spectrum into six centrality classes. The vertical lines indicate cuts which are used to define the centrality classes 2 - 6. Due to the low event statistics in the most central class, we employ a different data set with an online trigger on the 7\% most central interactions, thus increasing the statistics for this event class by a factor of 20. For details see table~\ref{table}.

\begin{table}
\caption{\label{table} Number of events $N_{evt}$ in each centrality class, centrality limits expressed by the fractions of the total cross section $\sigma/\sigma_{tot}$, and the mean number of wounded nucleons $\langle N_{W} \rangle$ derived using the Glauber model}
\begin{indented}
\item[]\begin{tabular}{@{}crcc}
\br
Class & $N_{evt}$ & $\sigma/\sigma_{tot} (\%)$ & $\langle N_{W} \rangle (Glauber)$\\
\mr
1   & 577605 &    0 - 7    & 349 \\
2   &  45151 &    5 - 12.5 & 281 \\
3   &  67241 & 12.5 - 23.5 & 204 \\
4   &  64354 & 23.5 - 33.5 & 134 \\
5   &  63279 & 33.5 - 43.5 &  88 \\
6   & 117506 &    $>$ 43.5 &  42 \\ 
\br
\end{tabular}
\end{indented}
\end{table}

\subsection{$\Xi$ Analysis}
\begin{figure}[h]
\begin{center}
\includegraphics[height=4.2cm, width=7.5cm]{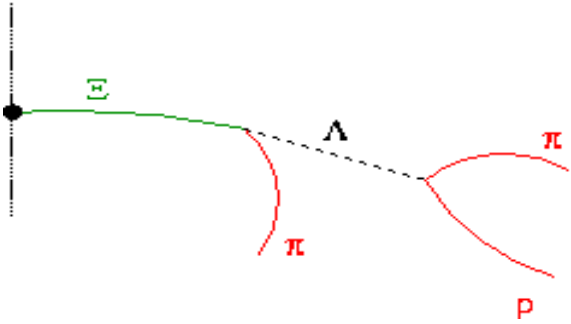}
\caption{\label{xi-decay} A sketch of the $\Xi$ decay topology}
\end{center}
\end{figure}

The $\Xi^{-} (\bar{\Xi}^{+})$ hyperon is not measured directly. It is identified on a statistical basis via its weak decay into $\Lambda (\bar{\Lambda})$ and $\pi^{-} (\pi^{+})$ (figure~\ref{xi-decay}). In order to find such decays, we first search for charged $\Lambda (\bar{\Lambda})$ decays by combining all negative tracks with positive ones. If a negative and a positive track have a distance of closest approach (DCA) less than 1~cm at any point along their trajectories, the pair is saved as a candidate $\Lambda (\bar{\Lambda})$. The contribution of false $\Lambda (\bar{\Lambda})$ candidates is reduced by requiring an energy loss in the TPCs of the positive (negative) particle according to the proton mass hypothesis. From these candidates we select those which are found in an invariant mass window $\pm$15MeV/$c^{2}$ around the $\Lambda$ mass \cite{PDB}. These candidates are combined with all negative (positive) tracks and if the DCA fell within 1 cm, the pair was saved as  $\Xi^{-} (\bar{\Xi}^{+})$ candidate. We use geometrical cuts to reduce the combinatorial background from uncorrelated $\Lambda (\bar{\Lambda})$ candidates and pions. By extrapolation of all particle trajectories backwards to the target plane, their impact parameters with respect to the primary interaction point are determined. We require the $\Xi$ to originate from and the $\Lambda$ to miss the primary interaction point. A further cut is applied to the decay position of the $\Xi$.

\begin{figure}[h]
\begin{center}
\includegraphics[height=6.0cm, width=9cm]{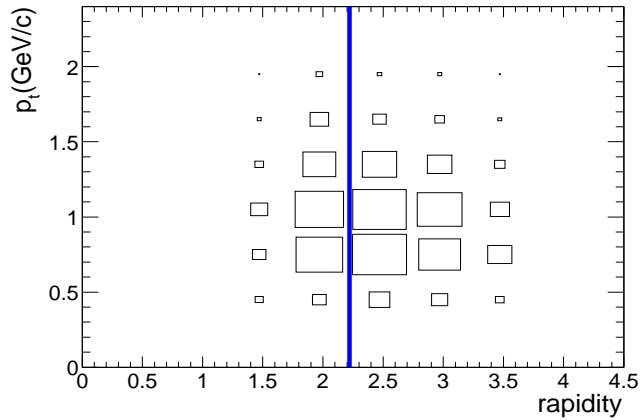}
\caption{\label{phasespace} Distribution of detected $\Xi^{-}$ in rapidity-$p_{t}$ space in central Pb+Pb collisions at 40~$A$\hspace{0.08cm}GeV. The line marks midrapidity (2.22).}
\end{center}
\end{figure}

\begin{figure}[h]
\begin{center}
\includegraphics[scale=0.82]{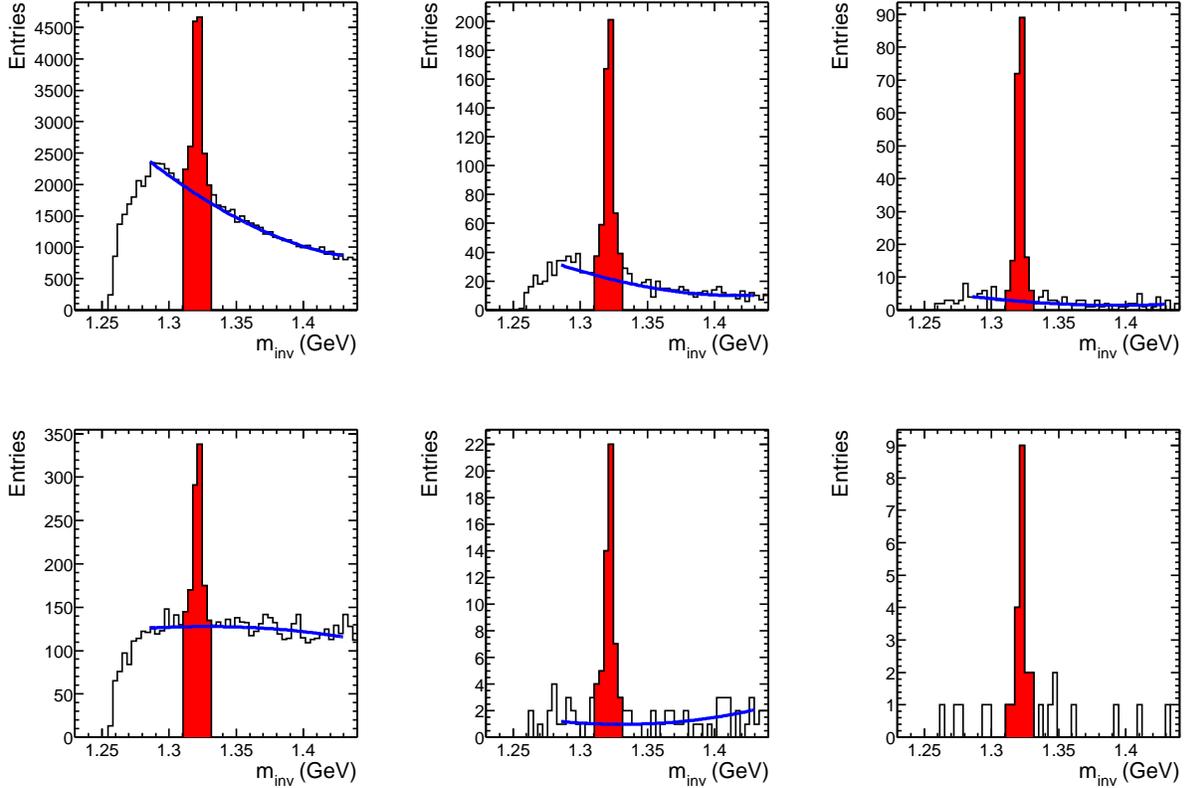}
\caption{\label{Minv} Invariant mass distributions for $\Xi^{-}$ (top) and $\bar\Xi^{+}$ (bottom) candidates at 40~$A$\hspace{0.08cm}GeV for three centrality classes (left: central(1), middle: midcentral(4), right: peripheral(6))}
\end{center}
\end{figure}

With this method we measure the $\Xi$ hyperons in a broad rapidity and transverse momentum range, see figure~\ref{phasespace}. However, for the following preliminary results on the $\bar{\Xi}^{+}/\Xi^{-}$ ratio we restrict ourselves to the midrapidity region $|y-y_{CM}|<0.5 $  ($y_{CM}=2.22 $) and to $p_{t}$ higher than 0.3 GeV/$c$. In figure~\ref{Minv} the corresponding invariant mass spectra of $\Xi^{-}$ and $\bar{\Xi}^{+}$ candidates for three centrality classes are shown. The combinatorial background increases with multiplicity and thus with centrality. By integrating the signal in the mass window $\pm$ 8 MeV/$c^{2}$ around the $\Xi$ mass ($m_{\Xi}=1.321$ GeV/$c^{2}$) \cite{PDB} and subtracting the combinatorial background obtained from a polynomial fit, we extract the raw number of $\Xi^{-}(\bar{\Xi}^{+})$. Systematic errors of the results presented in the following are not yet available.

\section{Results}

\begin{figure}[h]
\begin{center}
\includegraphics[scale=0.76]{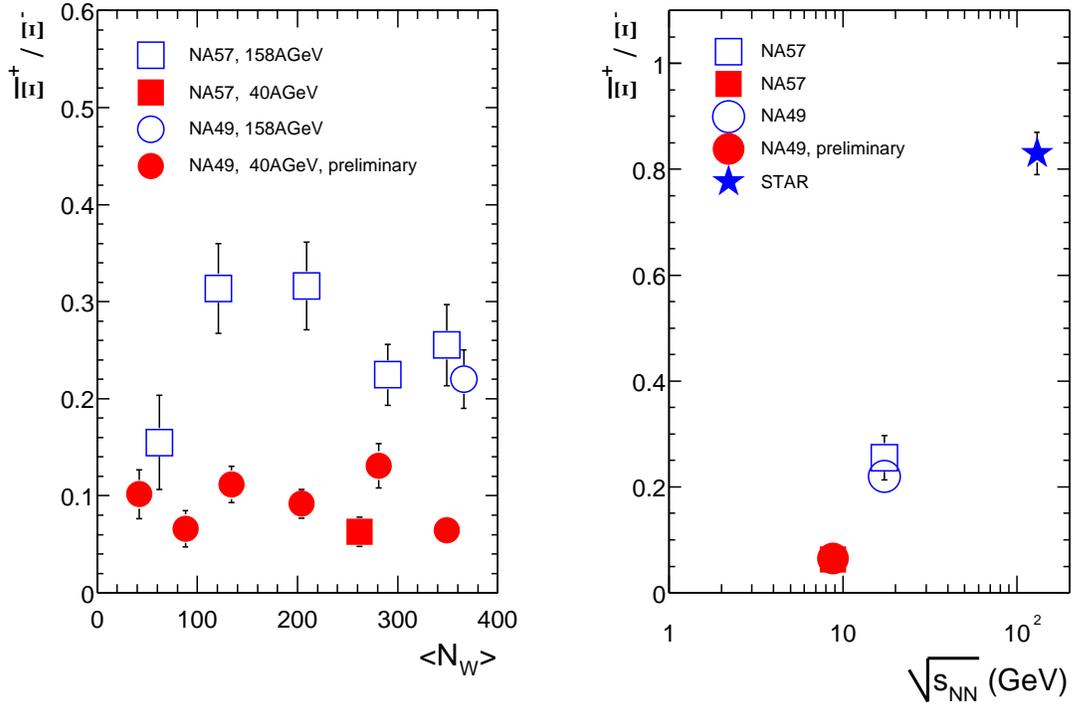}
\caption{\label{ratio}Left: Centrality dependence of the $\bar{\Xi}^{+}/\Xi^{-}$ ratio for Pb+Pb collisions at 40~$A$\hspace{0.08cm}GeV and 158~$A$\hspace{0.08cm}GeV. Right: Energy dependence of the $\bar{\Xi}^{+}/\Xi^{-}$ ratio. All ratios are at midrapidity and for central collisions. The errors are statistical only.}
\end{center}
\end{figure}

The centrality dependence of the $\bar{\Xi}^{+}/\Xi^{-}$ ratio is shown in figure~\ref{ratio} (left) together with corresponding results of NA49 \cite{Rob} and NA57 \cite{NA57} at 158~$A$\hspace{0.08cm}GeV. We note that the ratio is calculated using uncorrected data. However, the acceptance and detection efficiency for $\Xi^{-}$ and $\bar{\Xi}^{+}$ are expected to be similar and thus the corresponding corrections should approximately cancel in the ratio. This assumption is justified, since the transverse distributions of $\Xi^{-}$ and $\bar{\Xi}^{+}$ are similar, at least as measured at 158~$A$\hspace{0.08cm}GeV \cite{Rob}, and the longitudinal distributions around midrapidity do not strongly deviate. We observe no significant centrality dependence for the ratio at 40~$A$\hspace{0.08cm}GeV which is similar to what was reported by the NA57 collaboration at 158~$A$\hspace{0.08cm}GeV. The ratio at 40~$A$\hspace{0.08cm}GeV is smaller than the ratio at 158~$A$\hspace{0.08cm}GeV for all centrality classes. At both energies NA49 and NA57 data are consistent. \\
The energy dependence of $\bar{\Xi}^{+}/\Xi^{-}$ ratio for central Pb+Pb (Au+Au) collisions is shown in figure~\ref{ratio} (right), where also the results of NA49 \cite{Rob}, NA57 \cite{NA57} at 158~$A$\hspace{0.08cm}GeV and STAR \cite{STAR} at $\sqrt{s_{NN}}=130$ GeV are presented.  The $\bar{\Xi}^{+}/\Xi^{-}$ ratio increases strongly with the collision energy. This trend is observed for all other $\bar{B}/B$ ratios \cite{michi} and, within the statistical approach, is described by the decrease of the baryochemical potential $\mu_{B}$ with the collision energy. \\
Finally we compare the measured $\bar{\Xi}^{+}/\Xi^{-}$ ratio ($0.065 \pm  0.006$) at 40~$A$\hspace{0.08cm}GeV for central collisions with the hadron-gas model [1] with partial strangeness saturation. Based on the fit to several particle abundances measured in central Pb+Pb collisions at 40~$A$\hspace{0.08cm}GeV, excluding the $\Xi^{-}$ and $\bar{\Xi}^{+}$, it predicts the $\bar{\Xi}^{+}/\Xi^{-}$ ratio in 4$\pi$ to be 0.06 ($T=(147.7 \pm 2.0)$MeV, $\mu_{B}=(377.2 \pm 8.6)$MeV, $\gamma_{S}=0.747 \pm 0.023$). The measured ratio agrees nicely with the model prediction. Note, however, that the measurement is made at midrapidity, while the model prediction is for 4$\pi$ integrated values, and the ratio is calculated from yields not corrected for acceptance and efficiency.

\section{Summary and Outlook}
\begin{figure}[h]
\begin{center}
\includegraphics[scale=0.41]{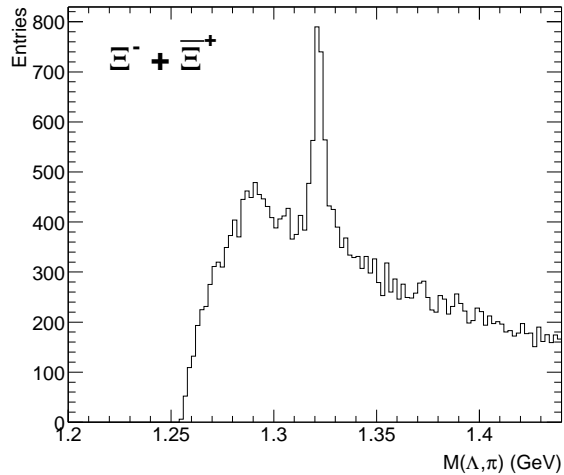} 
\caption{\label{Minv30} Invariant mass distribution for $\Xi^{-}+\bar{\Xi}^{+}$ candidates at 30~$A$~\hspace{0.08cm}GeV}
\end{center}
\end{figure}

We report the first preliminary results on the centrality dependence of the $\bar{\Xi}^{+}/\Xi^{-}$ ratio in Pb+Pb collisions at 40~$A$\hspace{0.08cm}GeV from NA49. We find no significant centrality dependence of this ratio. On the other hand the ratio increases strongly with the collision energy. Our result at 40~$A$\hspace{0.08cm}GeV is in good agreement with the hadron-gas model prediction. \\
In the near future we expect to obtain rapidity, $p_{t}$ spectra, and 4$\pi$ yields for $\Xi$ hyperons in different centrality classes at 40~$A$\hspace{0.08cm}GeV. In addition we are looking forward to results on $\Xi^{-}$ and $\bar{\Xi}^{+}$ production in central Pb+Pb collisions at 20, 30, and 80~$A$\hspace{0.08cm}GeV. Figure~\ref{Minv30} shows for the first time the invariant mass spectrum of $\Xi^{-}+\bar{\Xi}^{+}$ candidates at 30~$A$~\hspace{0.08cm}GeV.

\section*{Acknowledgments}
This work was supported by the Director, Office of Energy Research, 
Division of Nuclear Physics of the Office of High Energy and Nuclear Physics 
of the US Department of Energy (DE-ACO3-76SFOOO98 and DE-FG02-91ER40609), 
the US National Science Foundation, 
the Bundesministerium fur Bildung und Forschung, Germany, 
the Alexander von Humboldt Foundation, 
the UK Engineering and Physical Sciences Research Council, 
the Polish State Committee for Scientific Research (2 P03B 130 23, SPB/CERN/P-03/Dz 446/2002-2004, 2 P03B 02418, 2 P03B 04123), 
the Hungarian Scientific Research Foundation (T032648, T14920 and T32293),
Hungarian National Science Foundation, OTKA, (F034707),
the EC Marie Curie Foundation,
and the Polish-German Foundation.


\section*{References}
\begin{thebibliography}{20}

\bibitem{bgs98} Becattini F, Gazdzicki M, Sollfrank J 1998 {\it Eur. Phys. J.} C5 {\bf 143} and Becattini F {\it private communication}
\bibitem{NA49setup} Afanasiev S V \etal [NA49 collaboration] 1999 {\it Nucl. Instrum. Methods} A {\bf 430} 210
\bibitem{PDB} Hagiwara K \etal [Particle Data Group] 2002 {\it Phys. Rev.} D {\bf 66} 010001
\bibitem{Rob} Afanasiev S V \etal [NA49 collaboration] 2002 {\it Phys. Lett.} B {\bf 538} 275-281
\bibitem{NA57} Elia D for the NA57 collaboration 2002, to be published in {\it Nucl. Phys.} A (QM02 proceedings)
\bibitem{STAR} Castillo J for the STAR collaboration 2002, to be published in {\it Nucl. Phys.} A (QM02 proceedings)
\bibitem{michi} Mitrovski M for the NA49 collaboration {\it These proceedings}

\end{thebibliography}
\end{document}